\documentclass{PoS}

\title{Heavy-flavour production in Pb-Pb collisions at \\$\sqrt{s_{NN}}$ = 2.76 TeV with the ALICE detector}

\ShortTitle{Heavy-flavour production}

\author{\speaker{Alessandro Grelli for the ALICE Collaboration}\thanks{A list of members of the ALICE Collaboration and acknowledgments can be found at the end of this issue.}\\
        Utrecht University\\
        E-mail: \email{a.grelli@uu.nl}}

\abstract{Hadrons containing heavy-flavours, i.e. charm and beauty quarks, are unique probes of the properties of the hot and dense QCD medium produced in heavy-ion collisions. Due to their large masses, heavy quarks are produced at the
initial stage of the collision, almost exclusively via hard partonic scattering processes. Therefore, they are expected to experience the full collision history propagating through and interacting with the QCD
medium. The parton energy loss, which is sensitive to the transport coefficients of the produced medium, can be studied experimentally by measuring the nuclear modication factor ($R_{\rm AA}$) which accounts for
the modication of the heavy-flavoured hadron yield in Pb-Pb collisions with
respect to pp collisions. In semi-central Pb-Pb collisions, the degree of thermalization of charm quarks in the QCD medium can be accessed via the measurement of the heavy flavour elliptic flow $v_2$ at low $p_{\rm T}$ . At high $p_{\rm T}$, $v_2$
is sensitive to the path-length dependence of heavy quark in-medium energy loss. The ALICE collaboration has measured the production of open heavy  flavour hadrons via their hadronic and semi-leptonic decays at mid-rapidity and in the semi-muonic decay channel at forward rapidity in pp, p-Pb and Pb-Pb collisions at 7, 5.02 and 2.76 TeV respectively.
In this talk the current results on open heavy-flavour $R_{\rm AA}$ and $v_2$ will be presented.}

\FullConference{The European Physical Society Conference on High Energy Physics -EPS-HEP2013\\
		18-24 July 2013\\
		Stockholm, Sweden}

\begin{document}

\section{Introduction}

The main goal of the ALICE \cite{det} experiment is to study strongly interacting matter in the conditions 
of high density and temperature produced by colliding ultra-relativistic lead ions at the LHC. In such
 conditions lattice QCD calculations predict quark deconfinement and the formation of the so called Quark-Gluon Plasma (QGP) \cite{wil}. Heavy-flavour hadrons, containing
charm and beauty quarks are abundantly produced at LHC energies. They are regarded as effective probes of QGP thermodynamics and the medium's transport properties. Open charm and beauty hadrons should be sensitive to the energy density, through the mechanism of in-medium parton energy loss. The nuclear modification factor $R_{\rm AA}$ of particles
is well-established as a 
 sensitive observable for the study of the interaction of hard partons with the medium.
This factor is defined as the ratio of the $p_{\rm T}$-differential yield measured in nucleus-nucleus (AA) collisions in a given centrality class to the yield calculated from the proton-proton cross-section scaled by the nuclear overlap function $<T_{AA}>$ for that centrality class, obtained from Glauber model calculations of the collision geometry [3],
\begin{equation}
\label{eq:Raa}
R_{\rm AA}(p_T)=
{1\over <T_{\rm AA}>} \cdot 
{\mathrm{d} N_{\rm AA}/\mathrm{d}\ p_T \over 
\mathrm{d}\sigma_{\rm pp}/\mathrm{d}\ p_T}\,,
\end{equation}
A strong suppression of the yield of charged
particles was observed in Pb-Pb collisions at the LHC \cite{sup} relative to scaled pp collisions. Due to the QCD nature
of parton energy loss, quarks are predicted to lose less energy than gluons (that have
a higher color charge) and, in addition, the dead-cone effect and other mechanisms
are expected to reduce the energy loss of massive quarks with respect to light quarks \cite{dead}. Therefore, a hierarchy in the $R_{AA}$ is expected to be
observed when comparing the mostly gluon-originated
light-flavour hadrons (e.g. pions) to D and to B mesons [8]: 
$R^{\pi}_{\rm AA} < R^D_{\rm AA} < R^B_{\rm AA}$. 
The measurement and comparison of these different medium probes provides a unique test of
the color-charge and mass dependence of parton energy loss.
A full understanding of these phenomena requires also to quantify the initial-state effects inherent to nuclear collisions, like $k_{\rm T}$ broadening and nuclear PDF shadowing \cite{shad}. These effects are not related to the QGP phase but to cold nuclear matter and they may result in a modification of the $R_{\rm AA}$ shape (especially at low $
p_{\rm T}$), preventing to draw quantitative conclusions on the QGP properties. The initial-state effects are studied by measuring the modification of heavy-flavour hadron yields in proton-nucleus with respect to proton-proton collisions ($R_{\rm pA}(p_{\rm T})$).
Further insight in the medium properties is provided by the measurement of anisotropy in the
 azimuthal distribution of particle momenta.
 The azimuthal anisotropy of produced particles
 is characterized by the Fourier coefficients $v_n=<cos
[n( \phi-\Psi_n)]>
$, where $n$ is the order of the harmonic,
 $\phi$ is the azimuthal angle of the particle's momentum, and $\Psi_n$ is the azimuthal angle of the initial state symmetry plane for the $n$-th harmonic.
 At low ($p_T<3$ GeV/$c$) and intermediate (3-6 GeV/$c$) transverse momentum a non zero $v_2$  reflects the collective expansion, which is generated through interactions among the medium constituents. At high transverse momentum a positive $v_2$ is expected as a consequence of the path-length dependence of the in-medium parton energy loss. The charmed hadron $v_2$ offers a unique opportunity to test whether also quarks with large mass participate in the collective expansion dynamics and possibly thermalize in the QGP.
 
\section{Data sample}

A large Pb-Pb data sample at centre-of-mass energy $\sqrt{s_{\rm NN}}$ = 2.76 TeV was collected in November 2011. The events were triggered with a centrality-based selection using information from the VZERO scintillator arrays ($2.8 <\eta<5.1$ and $3.7 <\eta< 1.7$). Only events with a vertex found within 10 cm from the centre of the detector along the beam line were used, for a total of $16 \times10^6$ collisions in the 0-10$\%$ centrality class. About $4.5 \times10^6$ collisions were used in each of the centrality intervals: 10-20$\%$, 20-30$\%$, 30-40$\%$, 40-50$\%$ and 50-80$\%$. 
D mesons in the most peripheral centrality class and muons from heavy flavour decay were measured from the 2010 minimum-bias Pb-Pb sample for a total of about  $15 \times10^6$ events.  
The minimum-bias p-Pb data sample at $\sqrt{s_{\rm NN}}$ = 5.02 TeV, collected in January 2013, was triggered by requiring a signal in both the VZERO detectors. The efficiency of such a trigger for selecting non-single diffractive collisions is $>99\%$. The total p-Pb minimum-bias sample analyzed consists of about $12\times10^7$ events. 
 
\section{Open charm via D mesons}

The production of $D^0$, $D^+$, $D^{*+}$ \cite{7tev, PbPb} and $D^+_s$ was measured in pp, p-Pb and Pb-Pb collisions
at central rapidity ($|y| < 0.5$) via the exclusive reconstruction of the decays $D^0\rightarrow K^-\pi^+$, $D^+\rightarrow K^-\pi^+\pi^+$, $D^{*+}\rightarrow D^{0}\pi^+$ and $D^{+}_s\rightarrow \phi \pi^+\rightarrow K^-K^+\pi^+$. The analysis strategy for the extraction of the signal on top of a large combinatorial background is based on the reconstruction of
the D-meson decay topology. 
\begin{figure}[!htbp]  
\begin{center}        
\includegraphics[angle=0, width=6.65cm]{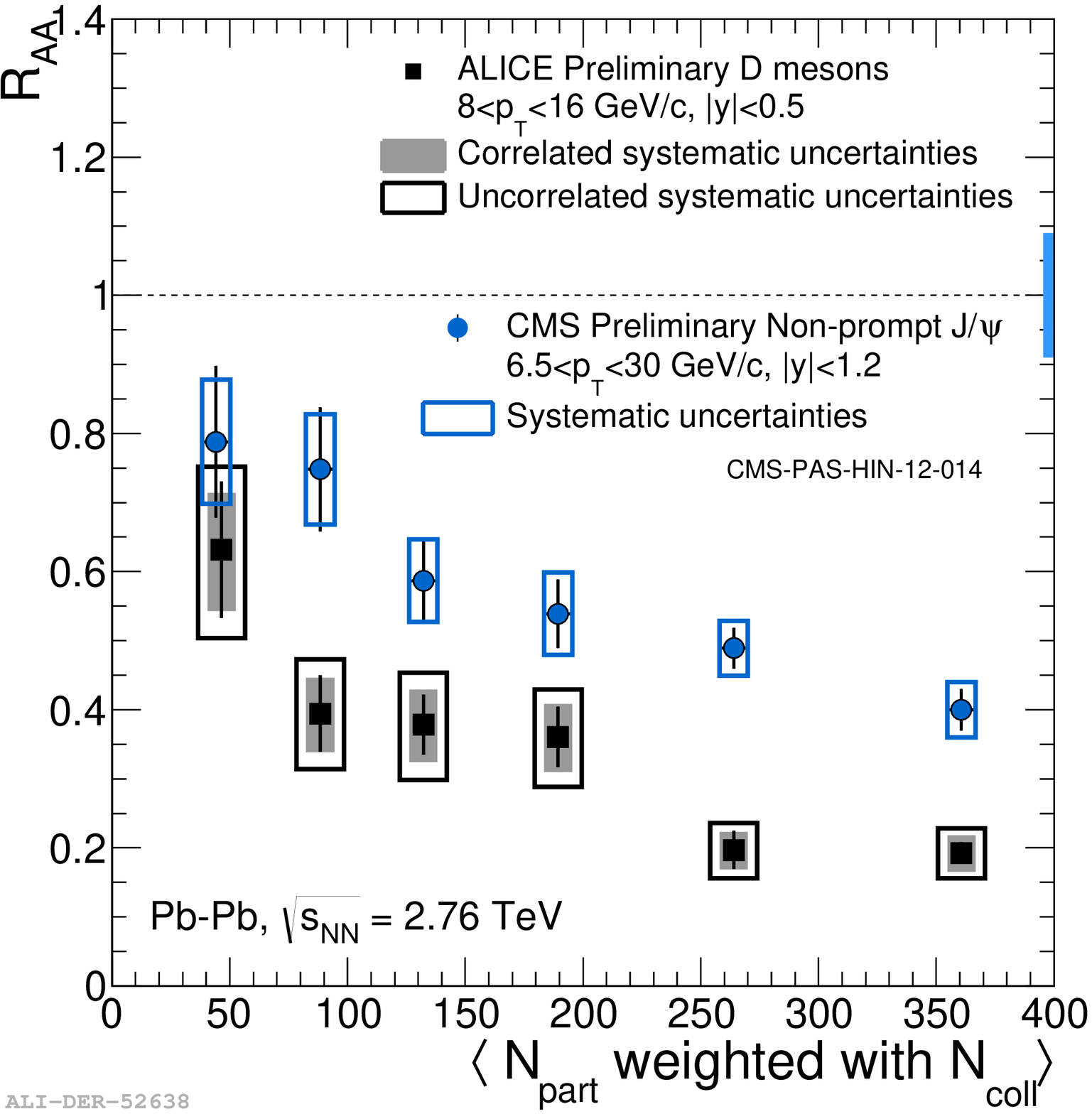}
\includegraphics[angle=0, width=7.05cm]{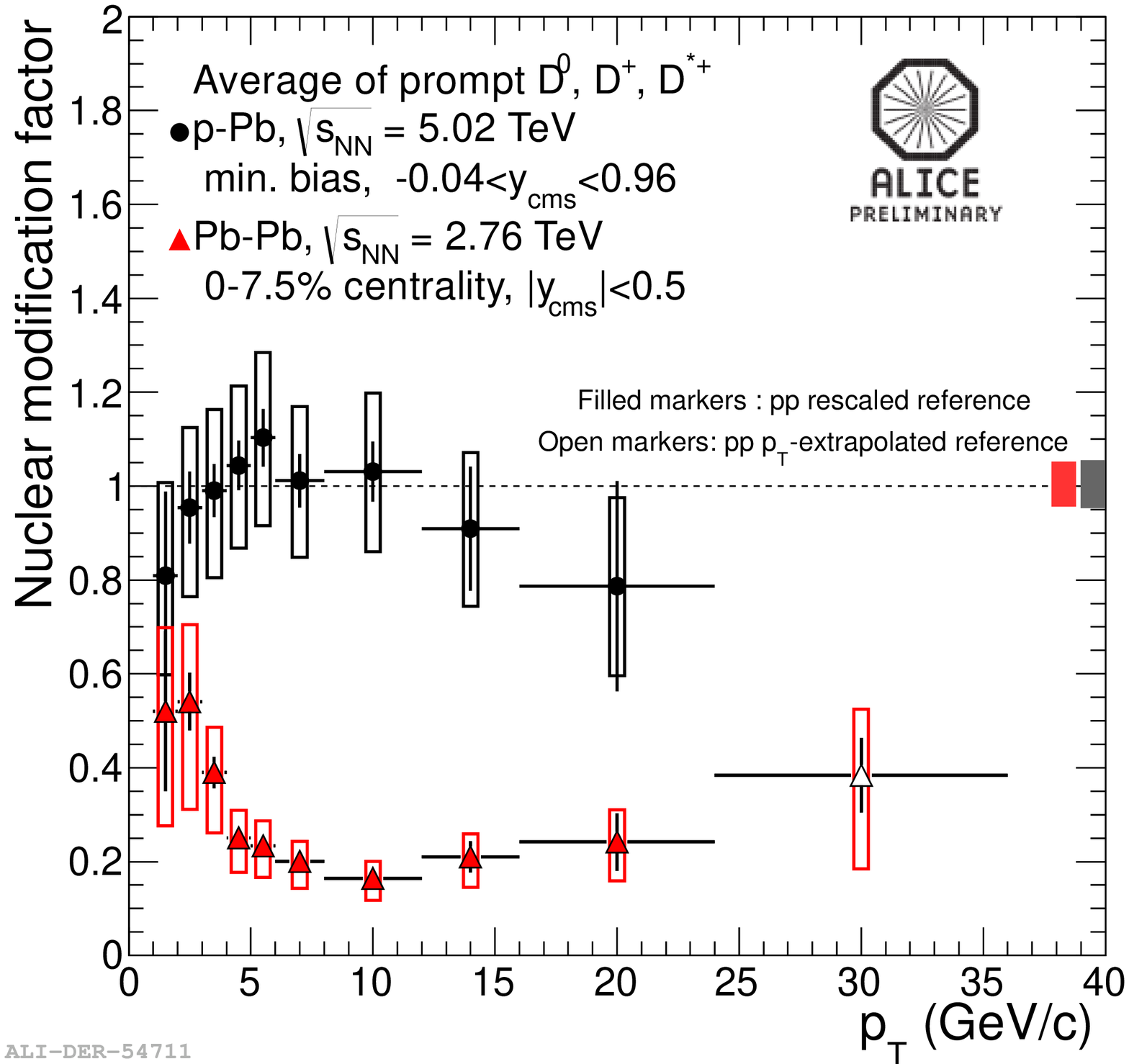}
\caption{ Left panel: Average $D^0$, $D^+$ and $D^{*+} $ $R_{\rm AA}$ as a function of centrality for $8<p_{\rm T}<16$ GeV/$c$, compared to non-prompt J/$\psi$ measured by the CMS Collaboration \cite{cms}.  Right panel: Average $p_{\rm T}$-differential D-meson $R_{\rm AA}$ in the centrality class 0-7.5$\%$ (red points) compared with centrality integrated $R_{\rm pPb}$ (black points).}
\label{fig:invmass}
\end{center}
\end{figure}
 The high resolution on the track position in the vicinity of the interaction point, provided by the Inner Tracking System (ITS) \cite{det}, allows us to resolve the  $D^0$, $D^+$ and $D^+_s$ decay vertices, which are typically displaced by few hundred micrometers from the primary vertex. D-meson candidates are selected by means of topological cuts. To further suppress the combinatorial background, particle identification (PID) on the D-meson decay products is employed. PID is performed using the information on specific energy deposit in the Time Projection Chamber (TPC) and on the velocity measured by the Time of Flight (TOF) detector.
  It allows a separation of kaons and pions in a wide momentum range up to 2 GeV/$c$. The signal yield is extracted by fitting the invariant mass distribution using a Gaussian function for the signal peak. 
The background is bescribed by an exponential shape in the case of $D^0$, $D^+$ while in the case
of the $D^{*+}$ a threshold function convoluted with an exponential is chosen. The correction for
efficiency and acceptance is performed using Monte Carlo simulations based on Pythia with Perugia-0
tuning and HIJING event generators. In order to extract the prompt D-meson yield, the contribution
of D mesons from B decays was evaluated relying on FONLL pQCD calculations \cite{fonll}. The reference pp measurement was obtained at $\sqrt{s}$ = 7 TeV and scaled to $\sqrt{s}$ = 2.76 TeV using FONLL.
The $R_{\rm AA}$ of prompt D mesons was measured in 10$\%$ wide centrality bins from 0 to 80$\%$ and in the following $p_{\rm T}$ regions: $2<p_{\rm T}<3$ GeV/$c$, $3<p_{\rm T}<5$ GeV/$c$, $5<p_{\rm T}<8$ GeV/$c$ and $8<p_{\rm T}<16$ GeV/$c$.
For $p_{\rm T}>$ 3 GeV/$c$ a suppression that increases going from peripheral to central collisions was found while in the region $2<p_{\rm T}<3$ GeV/$c$ the $D^0$ $R_{\rm AA}$ appear almost constant in all the centrality classes. The measurement, in the $p_T$ intervall $8<p_{\rm T}<16$ GeV/$c$ was compared with the $R_{\rm AA}$ of non-prompt $J/\Psi$ (from B-mesons decays)  measured by the CMS Collaboration \cite{cms}. 
As shown in the left panel of Fig.1, the comparison of the two results, in a similar kinematic range, shows clear indication for a mass hierarchy in the suppression  pattern being the non-prompt  $J/\psi$ less suppressed than D-mesons, as expected from a larger energy loss of charm quark with respect beauty quark. The right hand panel of Fig.1 shows the D-meson $R_{\rm pPb}$ (average of $D^0$, $D^+$ and $D^{*+}$). An $R_{\rm pPb}$ consistent with unity is observed with an hint of $R_{\rm pPb}<1$ for $p_{\rm T}<3$ GeV/$c$.
\begin{figure}[!htbp]  
\begin{center}        
\includegraphics[angle=0, width=6.7cm]{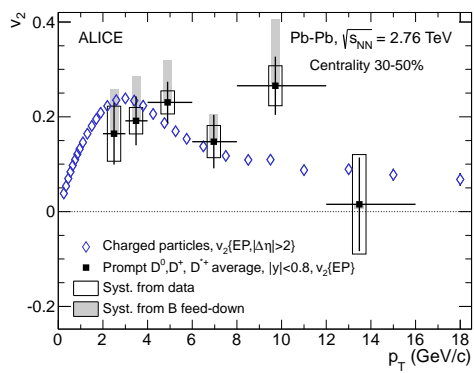}
\caption{ Average $D^0$, $D^+$ and $D^{*+} $ elliptic flow ($v_2$) as a function of transverse momentum in the Pb-Pb centrality interval 30-50$\%$ \cite{flow}.}
\label{fig:invmass2}
\end{center}
\end{figure}
The measurement of the D-meson elliptic flow, $v_2$, is shown in the Fig.2. The comparison with the charged particle $v_2$ shows a similar trend, within the rather large systematic and statistical uncertainties. The measurement indicates, at $5\sigma$ confidence level, a positive $v_2$ for $2<p_{\rm T}<6$ GeV/$c$ \cite{flow}. This suggests that charm quarks participate in the collective flow of the expanding medium.

\section{Heavy-flavour decay electrons}

The semi-leptonic decay of D and B mesons, with a branching ratio of $\sim 10\%$, gives access to
the investigation of open charm and beauty production. The identification of the electrons is
provided by the particle-identification system in the ALICE central barrel, namely the
TPC, TOF, Transition Radiation Detector (TRD) and ElectroMagnetic Calorimeter (EMCal). 
\begin{figure}[!htbp]  
\begin{center}        
\includegraphics[angle=0, width=6.cm]{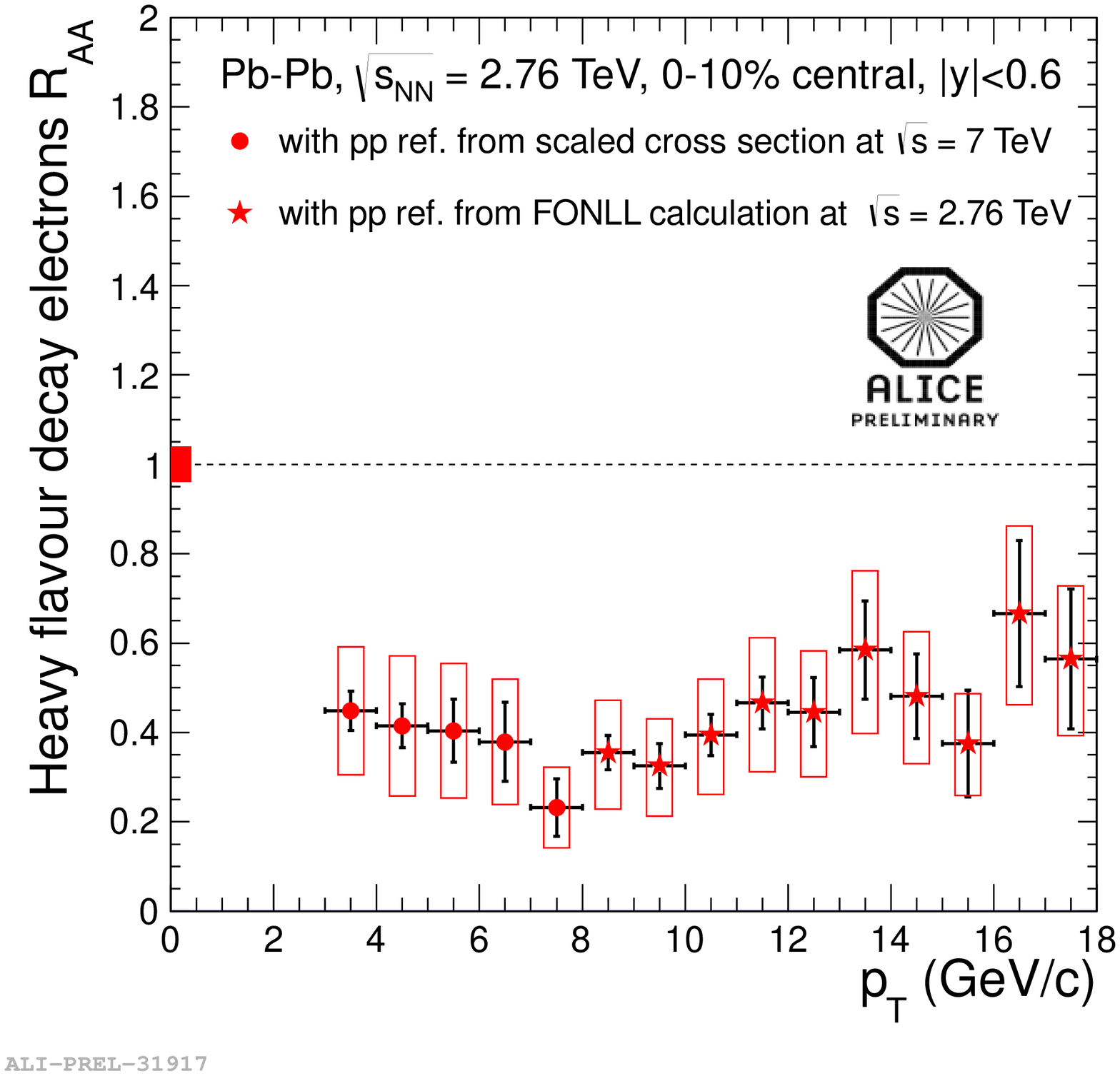}
\includegraphics[angle=0, width=8.0cm]{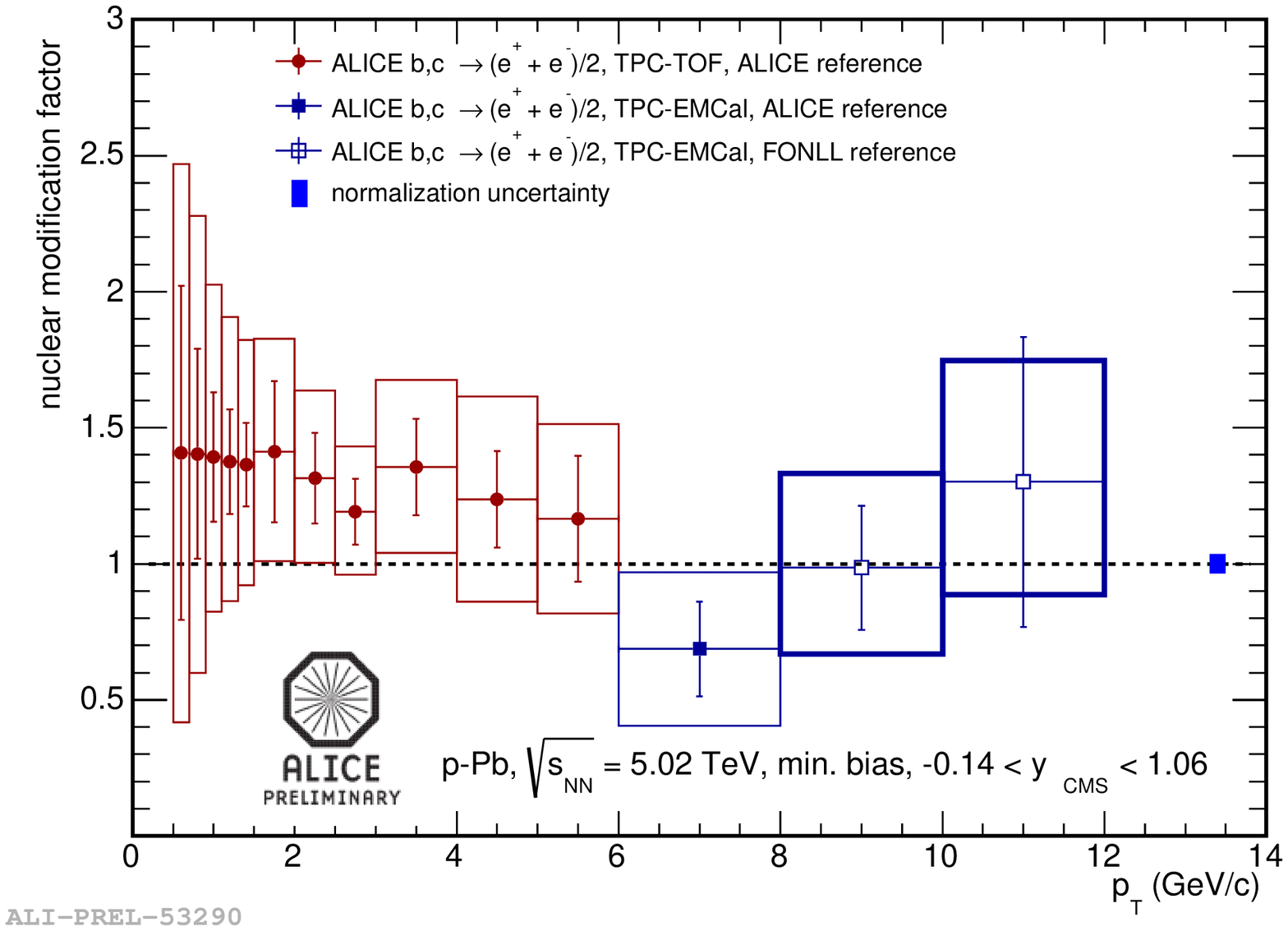}
\caption{ Left panel: Heavy-flavour decay electron $R_{\rm AA}$ as a function of $p_{\rm T}$ in the 10$\%$ most central collisions. Right panel: centrality integrated heavy-flavour decay electron $R_{\rm pPb}$.}
\label{fig:invmass3}
\end{center}
\end{figure}
TPC and EMCal were used for electron selection in Pb-Pb data. For the reference pp measurement, two different PID strategies were used. In addition to the common TPC PID, the first one employs the TOF and the TRD whereas the second one makes use of the EMCal. The reference measurement was obtained at $\sqrt{s_{NN}}$ = 7 TeV and then scaled to $\sqrt{s_{NN}}$ = 2.76 TeV using FONLL. 
The yield of electrons from heavy-flavour decays was obtained by subtracting background from other sources from the inclusive electrons yield. The $R_{\rm AA}$ of heavy-flavour decay electrons, and its dependence on centrality was measured with ALICE. In the left panel of Fig. 3 the $R_{\rm AA}(p_{\rm T})$ in the 10$\%$ most central events is presented. A suppression by a factor of 2-3 was found for $p_{\rm T}> 4$ GeV/$c$.  The nuclear modification factor $R_{pPb}(p_T)$, measured for centrality integrated p-Pb collisions, shown in the right panel of Fig. 3, is consitent with unity within substantial uncertanties.

\section{Single muons from heavy-flavour decay}

Muons originating from heavy-flavour hadron decays are measured with the forward muon spectrometer ($-4<\eta<-2.5$). Details on the detector and analysis strategy can be found in \cite{muons}. The heavy-flavour decay muon $R_{AA}$ measured as a function of the event centrality in the transverse momentum range $6<p_{\rm T}<12$ GeV/$c$ is shown in the left panel of Fig. 4. A suppression which increases from peripheral to central
collisions is observed. In the 10$\%$ most central collisions the suppression amounts to a factor of about 3, in agreement with the $R_{AA}$ measured
for heavy-flavour decay electrons at central rapidity.
\begin{figure}[!htbp]
\begin{center}        
\includegraphics[angle=0, width=6.8cm]{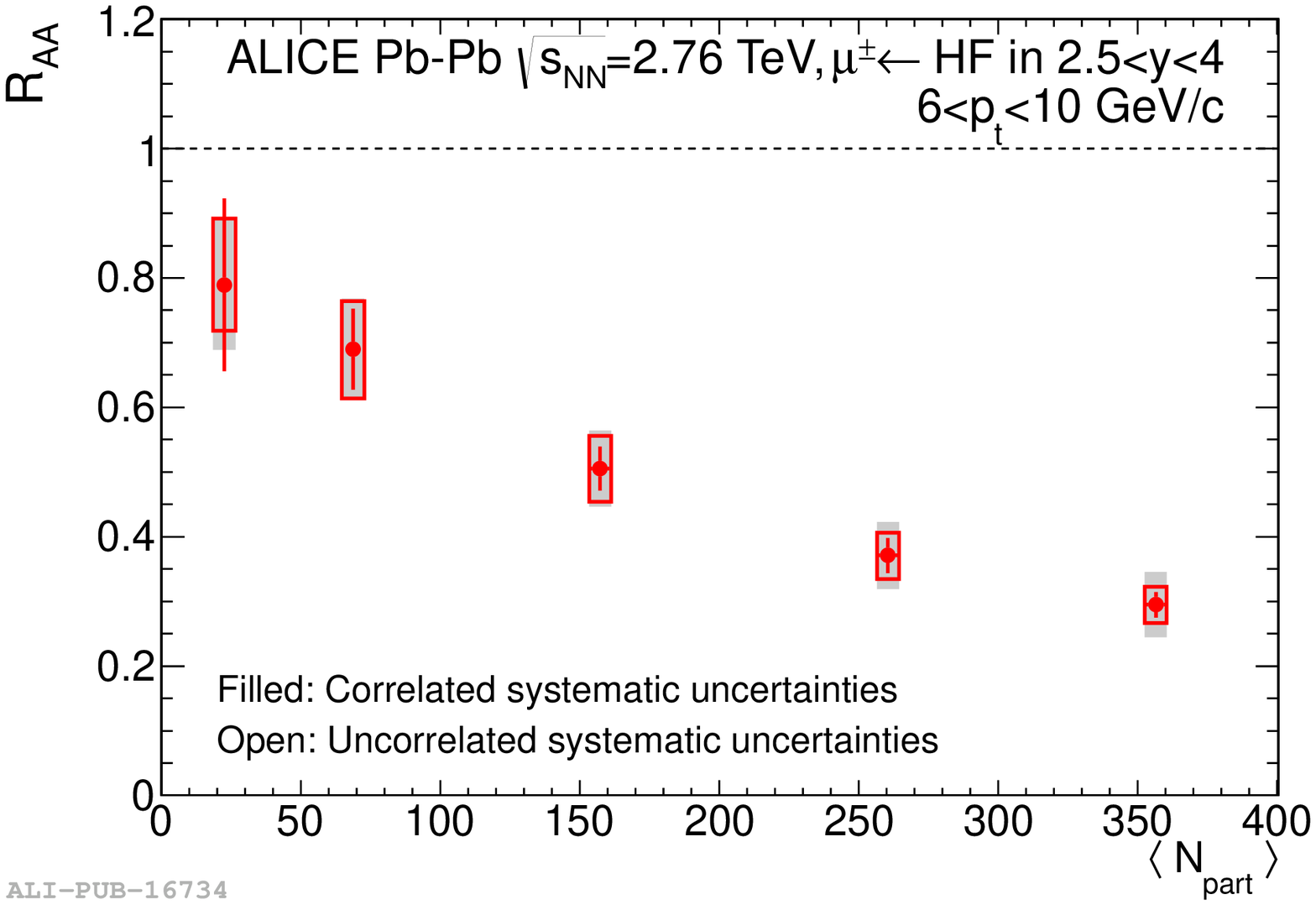}
\includegraphics[angle=0, width=6.8cm]{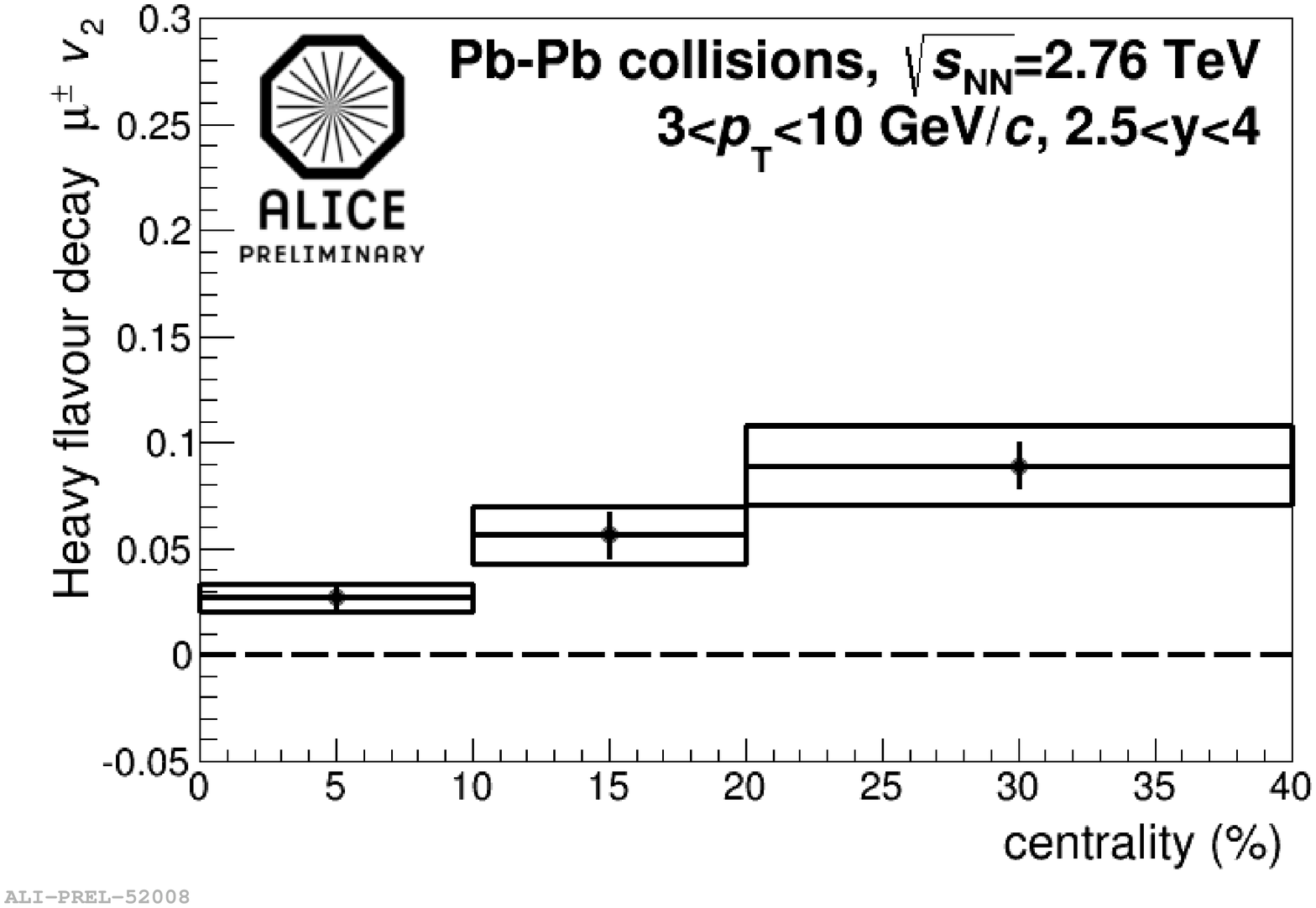}
\caption{ Left panel: Heavy-flavour decay muon $R_{\rm AA}$ as a function of centrality \cite{muons}. Right panel: Elliptic flow ($v_2$) of heavy-flavour decay muons as a function of the event centrality.}
\label{fig:invmass4}
\end{center}
\end{figure}
The measurement of the elliptic flow of muons from heavy-flavour hadron decays shows a positive $v_2$, consistently
with what observed for D mesons and heavy flavour decay electrons at central rapidity. In the right panel of Fig.4 the $v_2$ of muons shows a clear centrality dependence being larger for peripheral collisions.

\section{Conclusions}

In summary, the $R_{\rm AA}$ and $v_2$ of D mesons and heavy-flavour decay electrons at mid-rapidity and single muon from heavy-flavour decays at forward rapidity ($-4<|\eta|<-2.5$) were measured with ALICE. A positive elliptic flow of heavy-flavour hadrons was measured with a 5$\sigma$ significance in case of D mesons. The centrality dependent D-meson $R_{\rm AA}$, measured in the transverse momentum range $8<p_T<16$ GeV/$c$,
 was compared with the $R_{\rm AA}$ of non-prompt $J/\Psi$ \cite{cms} and a clear indication of a different suppression was found in agreement with the expected quark mass dependence of the energy loss. Finally, a measurements of the nuclear modification factor of
heavy-flavour hadron production in p-Pb collisions at $\sqrt{s_{NN}}$ = 5.02 TeV was presented. The results show an $R_{\rm pPb}$ compatible with unity in the measured transverse momentum range. This confirms that initial-state effects are small at high $p_T$ where a large suppression of the heavy-flavour hadron yield in Pb-Pb is observed, confirming that this suppression is due to a hot medium effect.

\end{document}